# Rapid IRMPD analysis for glycomics

Oznur Yeni,[a] Baptiste Schindler,[a] Baptiste Moge[a] and Isabelle Compagnon*[a]

Infrared vibrational spectroscopy in the gas phase has emerged as a powerful tool to determine complex molecular structures with Angstrom accuracy. Among the different approaches IRMPD (InfraRed Multiple Photon Dissociation), which requires the use of an intense pulsed tunable laser in the IR domain, has been broadly applied to the study of complex (bio)molecules. Recently, it also emerged as a highly relevant approach for analytical purposes especially in the field of glycomics in which structural analysis is still a tremendous challenge. This opens the perspective to develop new analytical tools allowing for the determination of molecular structures with atomic precision, and to address advanced questions in the field. However, IRMPD experiments require either non commercial equipments and long acquisition time which limits the data output. Here we show that it is possible to improve the IRMPD performances by optimizing the combination between a LTQ XL mass spectrometer and a high repetition tunable laser Firefly. Two orders of magnitude are gained with this approach compared to usual experiments ultimately leading to a completely resolved spectrum acquired in less than one minute. These results open the way to many new applications in glycomics with the possibility to include IRMPD in complex analytical workflows.

## Introduction

Vibrational spectroscopy in the gas phase is popular among the tools of the Physical Chemist for its atomic resolution of molecular structure and interactions, and its flawless pairing with Quantum Chemistry simulations.[1–3] In particular, UV-IR double resonance spectroscopy and rotational spectroscopy have been successfully used to study glycans.[4] In the 2000s, the coupling of such gas phase spectroscopic approaches with Mass Spectrometry, including efficient production, manipulation, selection and trapping of ions offered limitless possibilities in the field of molecular spectroscopy.[5]

More recently, vibrational ion spectroscopy has shown a solid potential for further applications in Analytical Chemistry.[6] In particular, the unique sensitivity of the vibrational fingerprint of a molecule to isomeric structural variations appeared as a game changer in the field of glycomics.[7–9] These developments are essential for the study of glycans, a class of biomolecules where minute structural variations affect the biological functions,[10] but also a daunting task because of their structural diversity. Indeed, glycans are known for their high degree of structural complexity: a reducing hexasaccharide can exist in more than $10^{12}$ potential structures.[11] Consequently, polymers become rapidly intractable and the development of structural analysis protocols become compulsory.[12,13]

Three mass spectrometry-based approaches have recently been used to resolve glycan ions structures using different spectroscopic schemes.[14] Messenger-tagging IR spectroscopy is based on the formation of a complex between the ion of interest and a "messenger", generally a gas (He, $N_2$) at cold temperature. A dissociation is observed when the ions interact with IR laser pulse and the degree of detachment is monitored by way of mass spectrometry.[15] IR spectroscopy in Helium droplet consists of capturing ions in helium dropplets. The helium atoms evaporate upon absorption of a photon, which is monitored by mass spectrometry.[8] These methods require single-photon absorption, in contrast to IRMPD (InfraRed Multiple Photon Dissociation). For IRMPD, the absorption of multiple infrared photons is needed to induce the fragmentation of the ions, which are isolated in a trap. The IRMPD spectrum is then obtained by monitoring the photofragmentation yield, calculated from parent and fragments ions intensities.[16]

IRMPD is the most frugal of these approaches. On one hand, it only offers a limited spectroscopic resolution as compared to cryogenic techniques. The structural resolution - that is, the capability to differentiate between two isomers - is however sufficient for small glycan units, and larger systems can be resolved by the analysis of its fragments.[7] On the other hand, the implementation is minimal because it operates at room temperature and requires very little modification of a commercial mass spectrometer. This is especially true in the case of a linear ion trap design with off axis ion detectors, which only require the opening of an IR transparent window at the rear end of the instrument. As such, this spectroscopic approach has an excellent potential for integration in analytical workflows.[17,18] One drawback of IRMPD is that a tunable IR laser delivering a relatively high power is needed because the excitation process requires several photons in a short delay to induce observable fragmentation. Such performances were first offered by the free electron lasers CLIO and FELIX[19,20] and quickly followed by table top systems. Such systems are typically pulsed and operate at 5 or 10 Hz, which causes long acquisition times and the main limitation of its use for analytical applications. Today, the acquisition of a high quality IR fingerprint in the 2700-3700 $cm^{-1}$ range (corresponding to the CH, NH and OH molecular vibrations) takes 42 minutes on our current setup operating at 10 Hz and with more than 10 mJ per pulse. In an effort to integrate IRMPD with upstream liquid chromatographic separation, this time was reduced to 5-6 minutes.[18] In this context, a gain in the repetition rate of mid-IR tunable lasers would immediately echo in better analytical performances. Very recently, kHz technology became available for mid-IR tuneable lasers, however with only 1μJ/pulse. Kong et al. remarkably shown that IRMPD spectra can be obtained at such low energy per pulse.[21] In their configuration, the kHz laser is coupled with a FT-ICR mass spectrometer. This type of design typically requires seconds of irradiation when using a 10 Hz OPO or a FEL.[22] This results in relatively slow acquisition of IR fingerprints, which is not a critical parameter for physical chemistry applications but is too slow for analytical applications.

Here we show the possibility to perform IRMPD spectroscopy in 14 seconds, gaining two orders of magnitude compared to the traditional approach. To do so, we combined a high repetition rate IR tunable laser (M Squared Firefly, 150 kHz, 150 mW) with a linear ion trap (ThermoFinnigan LTQ XL). The performance of this setup is illustrated on protonated glucosamine: it is demonstrated that the acquisition time can be decreased from 42 minutes to 14 seconds while keeping spectroscopic performances compatible with analytical chemistry.

## Experimental

### High speed IRMPD spectroscopy

Our new apparatus is presented in Figure 1. The setup used to acquire rapid spectra consists on the combination of a mass spectrometer, a LTQ XL commercialized by ThermoFinnigan, and a kHz tunable infrared laser, a Firefly commercialized by M Squared Lasers. The mass spectrometer is equipped with an electrospray ion source, ion guiding optics and a linear ion trap (only the trap is represented in Fig.1). The laser has a high repetition rate (150 kHz) and a high output power (>80 mW), with tunability range between 2.5-3.7 μm. This spectral range is suitable for IR absorption resonances with OH, NH and CH modes targeted here. The linewidth of the laser is less than 10 cm$^{-1}$ and the pulse duration is less than 10 nanoseconds (FWHM).

The LTQ XL has been modified to allow the injection of the laser beam into the ion trap. This requires drilling a hole in the mechanical architecture of the spectrometer and placing a MgF$_2$ window which is transparent to the IR light. The laser beam is directly injected through this window to the ion trap using the mirrors M1 to M3. Note that the Firefly is very compact (14 x 6 x 3 inches) and offers the possibility of future integration in analytical workflows.

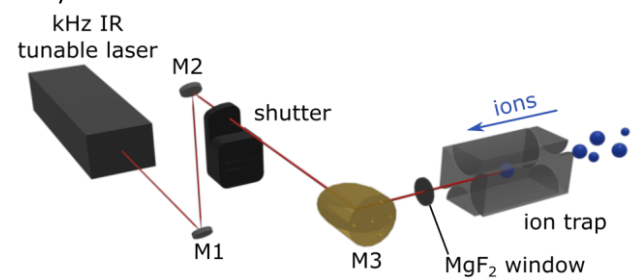

Figure 1: Optical coupling and synchronization of a kHz IR tunable laser with a linear ion trap. Focusing mirror M3 is used to optimize photofragmentation. A mechanical shutter is used for the synchronization.

A home-made software is used to control the laser (scanning range and speed) and to generate the wavelength data in parallel with the acquisition of the MS data by the commercial software. A mechanical shutter is used to synchronize the irradiation with the trapping time.

A typical irradiation time is 30 ms, which corresponds to 4500 laser shots during the trapping (Fig.2a). As a comparison, an IR spectrum recorded with the 10 Hz OPO laser uses an irradiation time of 700 ms corresponding to only 7 laser shots per trapping sequence (Fig.2b).

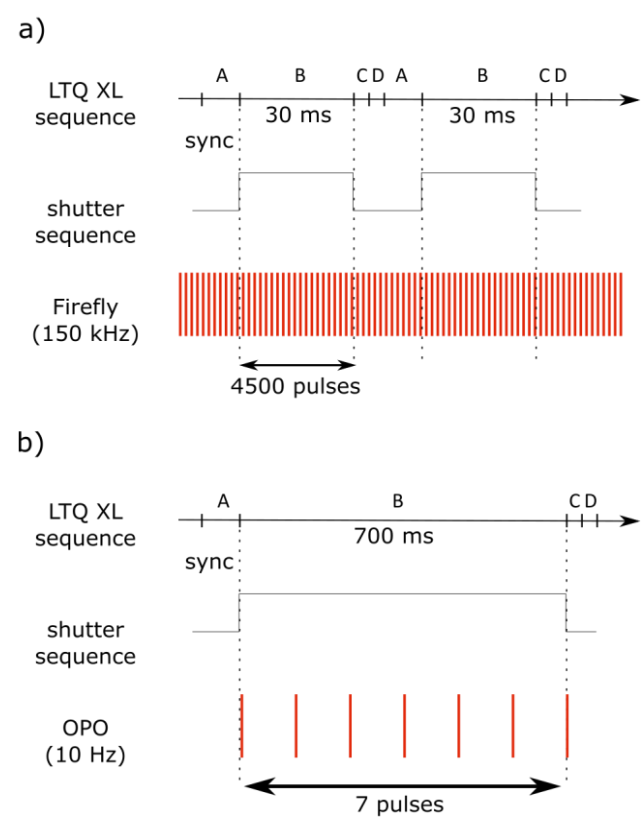

Figure 2: a) time sequence with the experimental setup LTQ XL – Firefly. b) time sequence with the experimental setup LTQ XL – OPO. With A= injection of ions in the trap, B= irradiation of ions of interest in the trap, C= ejection of ions from the trap and D= detection of ions.

In a typical IRMPD measurement, the variation of the fragmentation yield is recorded as a function of the IR wavelength. The corresponding absorption efficient is derived from this measurement following the formula:

$-\log \frac{I_p}{I_p + \sum I_f}$ with P the intensity of parent ion and F the summed intensity of each photofragment.

### Sample preparation

Glucosamine hydrochloride was purchased from Sigma. A solution of glucosamine was prepared at a concentration of 10 μM in water/methanol and 0.1% of acetic acid was added to promote protonation of glucosamine.

## Results

Protonated glucosamine was detected and isolated at m/z 180 and laser induced fragmentation was measured at 3345 cm$^{-1}$, which corresponds to one of the NH stretching frequency. The photofragmentation is detected by monitoring the m/z 162 fragmentation channel, which corresponds to the loss of a water molecule at the reducing end. The fragmentation ratio is calculated according to the spectroscopic convention (ratio of the intensity of fragment to the sum of the parent and fragment intensity) in contrast with the mass spectrometry convention (ratio of the fragment to the parent).

In the standard operating conditions of the 10 Hz laser (12mJ per pulse), the ions are irradiated for 700 ms to yield around 50% of photofragmentation. Using the kHz laser, the photofragmentation yield was surprisingly high and the irradiation time had to be reduced to 30 ms in order to obtain a similar photofragmentation with an output power of 144 mW at an operating repetition rate of 130 kHz.

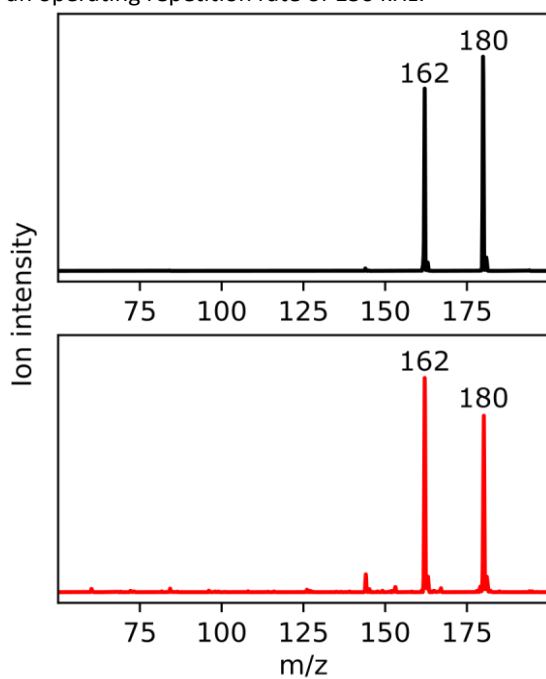

Figure 3: Photofragmentation spectra obtained at 3345 cm$^{-1}$ for protonated glucosamine with the setup up LTQ XL-Firefly (black) or LCQ-OPO (red)

The IRMPD spectrum of protonated glucosamine acquired in the 2700-3700 cm$^{-1}$ range using the conditions above with the kHz laser is compared to a reference IRMPD spectrum published previously using a Paul trap (LCQ ThermoFinnigan) and a 10 Hz OPO in Fig. 4.[7] First of all, vibrational bands are observed in the OH, NH and CH ranges in both cases and at the same frequency, resulting in overall similar spectra. Two main differences are observed between the two spectra: the relative intensity of some bands may vary. For example, the band at 3425 cm$^{-1}$ is less intense, relatively to the band at 3345 cm$^{-1}$, in Fig. 4a. Besides, the pattern of free OH vibrations present above 3600 cm$^{-1}$, which is partly resolved using the 10 Hz OPO, is entirely unresolved using the kHz OPO. This is consistent with the specifications of the kHz laser, which offers a 5 cm$^{-1}$ resolution in most of the spectral range but peaks at 10 cm$^{-1}$ around 3600 cm$^{-1}$. The former point is not unexpected: in multiple photon processes, the output is strongly energy dependant, and the smallest variation in the pulse intensity may result in significant changes in photofragmentation. As a consequence, band intensities may even vary from day to day on a given setup. Here, a weak band at 3050 cm$^{-1}$ in the reference spectrum is not visible in the new spectrum. Overall, the IRMPD spectrum of protonated glucosamine obtained using a kHz laser is highly recognisable when compared with the reference spectrum.

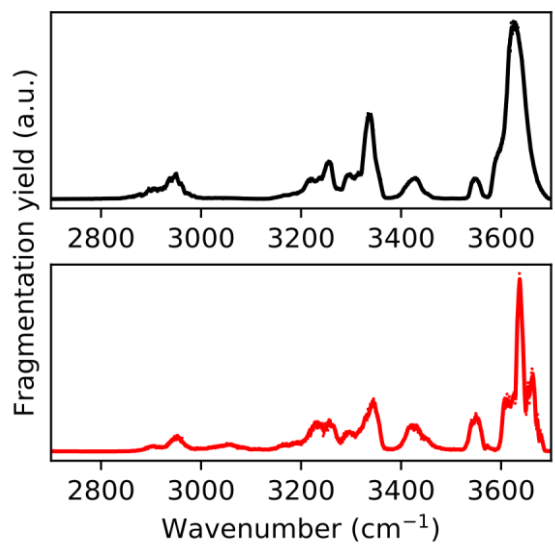

Figure 4: IRMPD spectra of protonated glucosamine acquired with the setup LTQ XL-Firefly (black) or LCQ-OPO (red)

Because the irradiation time was reduced from 700 ms to 30 ms by increasing the repetition rate of the laser, the acquisition time of the spectrum is reduced accordingly, from 42 to 18 minutes for 1700 data points (1.7 point per cm$^{-1}$). Moreover, because the vibrational features at room temperature are several cm$^{-1}$ broad, a reduction of the sampling should not result in a loss of information. This is confirmed in Fig. 5a and 5b: the sampling is reduced from 1700 points to 300 points by increasing the sweep speed setting from 0.9 cm$^{-1}$ per second to 20 cm$^{-1}$ per second and the spectrum remains visually identical. The following spectra (5c, 5d and 5e) were recorded when further reducing the sampling. At the maximum sweep speed of the laser, 75 datapoints are recorded in 14 seconds in the 2700-3700 cm$^{-1}$ range. While the detailed structure of the spectrum is overlooked at such speed, the main vibrational features are still present: the CH pattern around 2950 cm$^{-1}$; the two main features of the NH pattern at 3260 and 3345 cm$^{-1}$; and the three OH features at 3425, 3550 and 3640 cm$^{-1}$. This fingerprint is diagnostic of the glucosamine, which can be distinguished from the reference spectra of its isomers[7] in these experimental conditions.

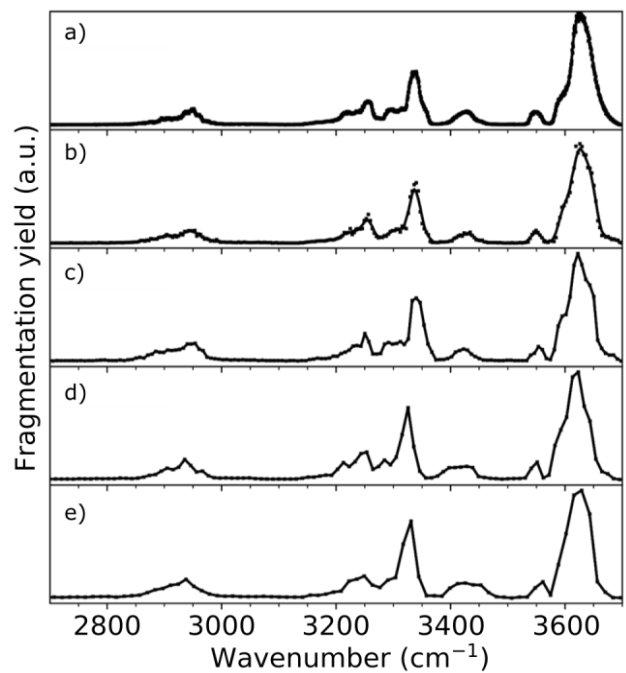

Figure 5: IRMPD spectra of protonated glucosamine acquired using the setup LTQ XL-Firefly with different times of acquisition: a) 18 minutes, b) one minute, c) 30 seconds, d) 20 seconds and e) 14 seconds

## Discussion

Photofragmentation induced by a laser with high repetition rate was surprisingly effective although only 1 µJ per pulse is available instead of 12mJ per pulse at 10 Hz. We obtained rapid IRMPD spectra of the same quality as reference spectra. At such low pulse energy, only protonated ions can be photofragmented and anions or metal cations complexes, which typically have higher fragmentation energy, are not accessible. Years ago, only protonated ions were accessible with available IR sources and a $CO_2$ laser was commonly used to enhance fragmentation.[22] Progress in the design of IRMPD setups has made possible the IRMPD spectra acquisition of anions and metal cations with a 10 Hz OPO laser.[7,23] Therefore, one can expect that the applicability of kHz OPOs for IRMPD application will expand in the coming years. However, we have recently reported that $NH_4^+$ can be used as an alternative for protonation, in the case of neutral molecules which do not readily protonate.[24] Such complexes are efficiently photofragmented for IRMPD spectroscopy using the kHz setup (Fig. S1).

Surprisingly, only 30 ms of irradiation were needed to yield 50% of photofragmentation of protonated glucosamine in the linear trap of the LTQ XL resulting in remarkable performance in terms of acquisition speed. We have reported that the quality of a 1 minute kHz spectrum is comparable to this of a 40 minutes reference spectrum; while a 14 seconds diagnostic signature could be obtained at the highest sweep speed of the laser. Note that we presented 1000 $cm^{-1}$ range fingerprints, while smaller ranges - hence shorter acquisition times - could be sufficient for analytical purposes. For example, the CH elongation modes are not generally isomer dependant, and are therefore of low utility for glycomics applications and could be left off the fingerprint. Considering the compactness of the laser system and the speed of fingerprint acquisition, one can envision the development of the integration of IRMPD spectroscopy in advanced analytical workflows. In particular, only seconds of ion signals are available for a diagnostic after chromatographic separation. We previously reported minutes long IRMPD acquisition coupled with stop-flow HPLC-MS.[18] The newly obtained performances are now readily compatible with HPLC-MS analysis.

In this work, we have compared the IR fingerprints of glucosamine in two significantly different experimental setups. In the classic setup, the ions are isolated in a Paul trap and irradiated by a few 12 mJ pulses. In the new setup, the ions are isolated in a linear trap and irradiated by thousands of 1µJ pulses. The fact that the fingerprints acquired on these two setups are similar demonstrates that IRMPD is reproducible, a strong argument for its use in analytical chemistry. Additionally, because the ions are only irradiated by a few pulses in the classic setup, the data are very sensitive to the pulse variations. In the new setup, such variations are averaged on thousands of pulses for each datapoint. This result in smother spectra as seen in Fig. 4 and in a higher confidence in the kHz data, which is another advantage for analytical applications.

## Conclusions

As a conclusion, a new IRMPD experimental setup was presented and consists in a combination between a LTQ XL mass spectrometer and a high repetition rate IR tunable laser. We have demonstrated that IRMPD spectra obtained with this instrument are similar in quality with ones acquired with our standard 10 Hz setup. The photofragmentation is remarkable efficient, which allows to obtain high speed IRMPD measurements, thus decreasing the time of acquisition for a spectrum in the range 2700 – 3700 $cm^{-1}$ from 42 minutes to one minute. At this speed, the spectrum is perfectly reproduced. The acquisition speed can be further optimized at the cost of resolution: in a spectrum acquired in only 14 seconds, the main features of vibrational absorption are clearly distinguishable. When comparison with quantum chemistry calculations is required, the spectral resolution must remain sufficient to provide enough information. For analytical purposes, rapid acquisition is often required and is attained with the reported performance. The compact laser technology and the automatized data acquisition offer the possibility of integration in complex analytical workflows.

## Author Contributions

O.Y.: Investigation, Visualization, Writing – original draft, Writing – review & editing. B.S.: Investigation, Software, Visualization, Writing – review & editing. B.M.: Writing – review & editing. I.C.: Conceptualization, Funding acquisition, Supervision, Writing – original draft, Writing – review & editing.

ORCID

O. Yeni: 0000-0001-6171-3096

B. Schindler: 0000-0002-7376-4154

B. Moge: 0000-0002-4932-6357

I. Compagnon: 0000-0003-2994-3961

## Conflicts of interest

There are no conflicts to declare.

## Acknowledgements

This work was supported by ANR ALGAIMS (ANR-18-CE29-0006-03, https://algaims-35.webself.net/accueil).

## Notes and references

1 J. P. Simons, *Molecular Physics*, 2009, **107**, 2435–2458.
2 S. Weiss, *Science*, 1999, **283**, 1676.
3 M. S. de Vries and P. Hobza, *Annual Review of Physical Chemistry*, 2007, **58**, 585–612.
4 E. J. Cocinero and P. Çarçabal, in *Gas-Phase IR Spectroscopy and Structure of Biological Molecules*, eds. A. M. Rijs and J. Oomens, Springer International Publishing, Cham, 2015, pp. 299–333.
5 H. Knorke, J. Langer, J. Oomens and O. Dopfer, *ApJ*, 2009, **706**, L66–L70.
6 R. E. van Outersterp, K. J. Houthuijs, G. Berden, U. F. Engelke, L. A. J. Kluijtmans, R. A. Wevers, K. L. M. Coene, J. Oomens and J. Martens, *International Journal of Mass Spectrometry*, 2019, **443**, 77–85.
7 B. Schindler, L. Barnes, G. Renois, C. Gray, S. Chambert, S. Fort, S. Flitsch, C. Loison, A.-R. Allouche and I. Compagnon, *Nat Commun*, 2017, **8**, 1–7.
8 E. Mucha, A. I. González Flórez, M. Marianski, D. A. Thomas, W. Hoffmann, W. B. Struwe, H. S. Hahm, S. Gewinner, W. Schöllkopf, P. H. Seeberger, G. von Helden and K. Pagel, *Angewandte Chemie International Edition*, 2017, **56**, 11248–11251.
9 C. Masellis, N. Khanal, M. Z. Kamrath, D. E. Clemmer and T. R. Rizzo, *J. Am. Soc. Mass Spectrom.*, 2017, **28**, 2217–2222.
10 A. Varki, *Glycobiology*, 2017, **27**, 3–49.
11 R. A. Laine, *Glycobiology*, 1994, **4**, 759–767.
12 R. D. Cummings and J. M. Pierce, *Chemistry & Biology*, 2014, **21**, 1–15.
13 C. J. Gray, L. G. Migas, P. E. Barran, K. Pagel, P. H. Seeberger, C. E. Eyers, G.-J. Boons, N. L. B. Pohl, I. Compagnon, G. Widmalm and S. L. Flitsch, *Journal of the American Chemical Society*, 2019, **141**, 14463–14479.
14 A. M. Rijs and J. Oomens, in *Gas-Phase IR Spectroscopy and Structure of Biological Molecules*, eds. A. M. Rijs and J. Oomens, Springer International Publishing, Cham, 2014, vol. 364, pp. 1–42.
15 N. Khanal, C. Masellis, M. Z. Kamrath, D. E. Clemmer and T. R. Rizzo, *Analytical Chemistry*, 2017, **89**, 7601–7606.
16 B. Schindler, J. Joshi, A.-R. Allouche, D. Simon, S. Chambert, V. Brites, M.-P. Gaigeot and I. Compagnon, *Phys. Chem. Chem. Phys.*, 2014, **16**, 22131–22138.
17 B. Schindler, A. D. Depland, G. Renois-Predelus, G. Karras, B. Concina, G. Celep, J. Maurelli, V. Loriot, E. Constant, R. Bredy, C. Bordas, F. Lépine and I. Compagnon, *International Journal for Ion Mobility Spectrometry*, 2017, **20**, 119–124.
18 B. Schindler, G. Laloy-Borgna, L. Barnes, A.-R. Allouche, E. Bouju, V. Dugas, C. Demesmay and I. Compagnon, *Anal. Chem.*, 2018, **90**, 11741–11745.
19 C. Kapota, J. Lemaire, P. Maître and G. Ohanessian, *J. Am. Chem. Soc.*, 2004, **126**, 1836–1842.
20 J. J. Valle, J. R. Eyler, J. Oomens, D. T. Moore, A. F. G. van der Meer, G. von Helden, G. Meijer, C. L. Hendrickson, A. G. Marshall and G. T. Blakney, *Review of Scientific Instruments*, 2005, **76**, 023103.
21 Y. Yang, G. Liao and X. Kong, *Sci Rep*, 2017, **7**, 16592.
22 J. K. Martens, I. Compagnon, E. Nicol, T. B. McMahon, C. Clavaguéra and G. Ohanessian, *J. Phys. Chem. Lett.*, 2012, **3**, 3320–3324.
23 B. Schindler, L. Barnes, C. J. Gray, S. Chambert, S. L. Flitsch, J. Oomens, R. Daniel, A. R. Allouche and I. Compagnon, *The Journal of Physical Chemistry A*, 2017, **121**, 2114–2120.
24 J. S. Ho, A. Gharbi, B. Schindler, O. Yeni, R. Brédy, L. Legentil, V. Ferrières, L. L. Kiessling and I. Compagnon, *J. Am. Chem. Soc.*, 2021, jacs.0c11919.